\documentclass[prd,superscriptaddress,reprint]{revtex4-1}

\usepackage{dcolumn}
\usepackage{bm}
\usepackage{graphicx}
\usepackage{lineno}

\usepackage{psfrag}
\usepackage{color}
\usepackage{hepunits}

\usepackage{graphicx,epsfig,amsfonts,amsthm}
\usepackage{subfigure}
\usepackage{amsmath}

\newcommand{\bk}{{\bf k}}

\newcommand{\bq}{{\bf q}}

\begin{document}


\newcommand{\sidebyside}[7]
{
\begin{figure*}[#1]
  \begin{minipage}[t]{0.47\linewidth}
    \centering
    \includegraphics{#2.eps}
    \caption{#4}
    \label{#3}
  \end{minipage}\hfill%
  \begin{minipage}[t]{0.47\linewidth}
    \centering
    \includegraphics{#5.eps}
    \caption{#7}
    \label{#6}
  \end{minipage}
\end{figure*}
}

\title{Uncertainties on exclusive diffractive Higgs and jets production at the
LHC}
\author{A. Dechambre}\affiliation{IFPA, Dept. AGO, Universit\'e de Li\`ege}\affiliation{CEA/IRFU/Service de physique des particules, CEA/Saclay}
\author{O. Kepka}\affiliation{Center for Particle Physics, Institute of Physics, Academy of Science, Prague}
\author{C. Royon}\affiliation{CEA/IRFU/Service de physique des particules, CEA/Saclay} 
\author{R. Staszewski}
\affiliation{Institute of Nuclear Physics, Polish Academy of Sciences, Krakow} 
\affiliation{CEA/IRFU/Service de physique des particules, CEA/Saclay} 

\begin{abstract} 
Two theoretical descriptions of exclusive diffractive jets and Higgs production
at the LHC were implemented into the FPMC generator: the Khoze, Martin, Ryskin
model and the Cudell, Hern\'andez, Ivanov, Dechambre exclusive model. We then
study the uncertainties. We compare their predictions to the CDF measurement
and discuss the possibility of constraining the exclusive Higgs production at
the LHC with early measurements of exclusive jets. We show that the present
theoretical uncertainties can be reduced with such data by a factor of 5. 
\end{abstract}

\maketitle

\section{Introduction} \label{sec:intro} 

The Higgs boson is the last particle of the Standard Model remaining to be
confirmed experimentally. Inclusive searches in decay channels such as $b\bar
b$, $W^+W^-$, $ZZ$, $\gamma\gamma$ and associated production have been
performed at the Tevatron and are being started at the LHC.  However the search
for the Higgs boson at low mass is complicated due to the huge background
coming from QCD jet events. Especially the $b\bar b$ channel, dominant for
$m_H=\unit{120}{\GeV}$, is very difficult at the Tevatron and literally
impossible at the LHC.  Thus other possibilities have been investigated, in
particular using the exclusive diffractive production
\cite{Khoze:2000cy,Boonekamp:2004nu}. In such processes both incoming hadrons,
$p\bar p$ at the Tevatron and $pp$ at the LHC, remain intact after the
interaction and the Higgs decays in the central region. The process involves
the exchange of a color singlet and large rapidity gaps remain between the
Higgs and the outgoing hadrons.  At the Tevatron it is not possible to produce
exclusively the Higgs boson due to the tiny cross section. However other
particles, or systems of particles, can be produced, \textit{i.e.} a pair of
jets (a dijet), $\chi_c$ or $\gamma\gamma$, as long as they have $0^{++}$
quantum numbers.

Since the incoming hadrons remain intact, lose a part of their energy and are
scattered at very small angles, it is experimentally possible to measure all
final state particles, including the scattered protons. This can be done using
detectors inserted close to the beam pipe at a large distance from the
interaction point.  Besides, at the Tevatron and for low luminosity at the LHC,
it is also possible to use the rapidity gap method to select such events.  A
big advantage of the exclusive production of the Higgs boson is a very accurate
mass determination from the measurement of the scattered proton energy
loss~\cite{Albrow:2000na,Staszewski:2009sw}.  In addition, if the Higgs is
observed in this mode at the LHC it ensures it is a $0^{++}$
particle~\cite{Khoze:2000cy}.

The plan of this paper is as follows. In section II we give an introduction to
the theoretical description of exclusive production and introduce two models:
the Khoze, Martin, Ryskin (KMR) and the Cudell, Hern\'andez, Ivanov, Dechambre
exclusive (CHIDe) model, and also discuss the sources of their uncertainties.
In section III the Forward Physics Monte Carlo (FPMC) program is presented and
the implementation of both models is discussed.  Section IV focuses on the CDF
measurement of exclusive jets production and shows that both models give
similar, reasonable descriptions of the data. In section V we analyze the
uncertainties using the CHIDe model as an example. Predictions for exclusive
production at the LHC are given in section VI, where in addition we study the
possibility of constraining the Higgs production at the LHC from early LHC
exclusive jets measurement. Finally, conclusions are given in section VII.

\section{Theoretical description} 

%
\begin{figure}[h!] \centering
\subfigure[]{\includegraphics[width=3cm]{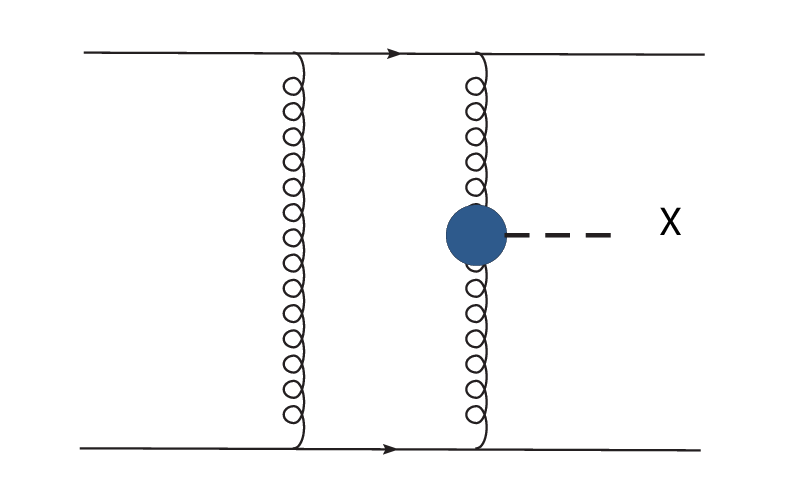}\label{fig_scheme_a}} \quad
\subfigure[]{\includegraphics[width=3cm]{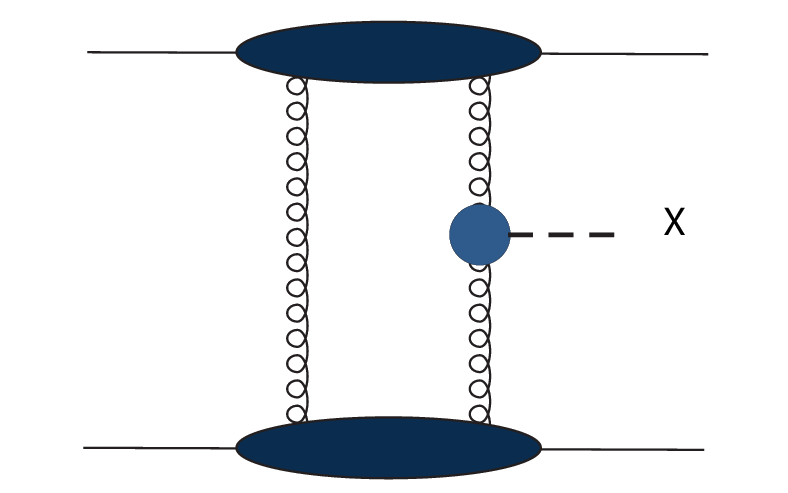}\label{fig_scheme_b}} \quad
\subfigure[]{\includegraphics[width=3cm]{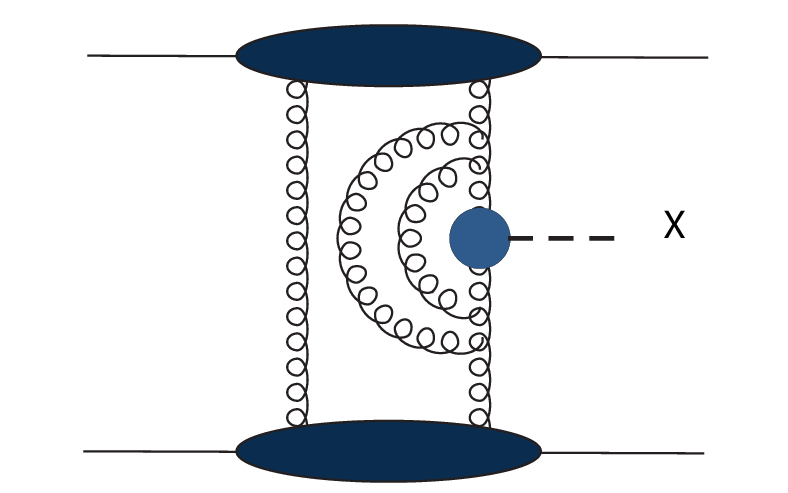}\label{fig_scheme_c}} \quad
\subfigure[]{\includegraphics[width=3cm]{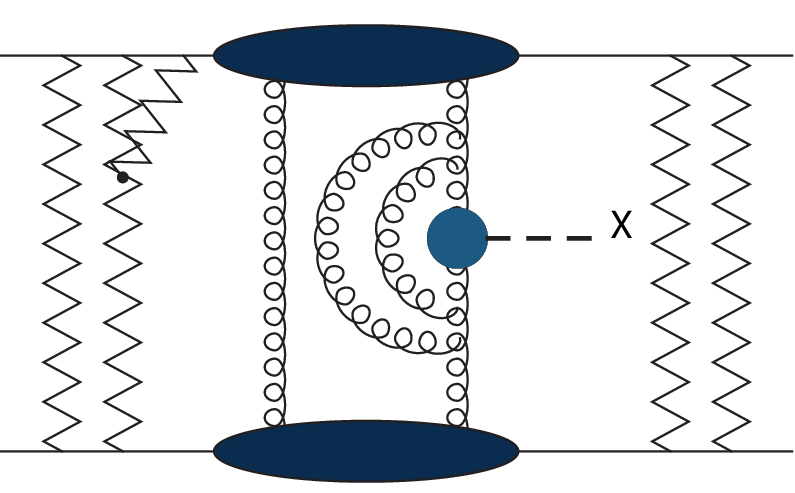}\label{fig_scheme_d}}
\caption{Schematic representation of the standard scheme of the exclusive cross
section calculation with its various steps. (a) Parton level calculation, (b)
impact factor, (c) Sudakov form factor and (d) rescattering
corrections.}\label{fig_scheme}  \end{figure}  
%

The exclusive production can be modeled in the QCD framework where the process
was described as a two-gluon exchange between quarks -- one gluon involved in
the production and the other one screening the color.  Such calculation
requires an analytic evaluation of a set of Feynman diagrams that describe the
production of a color-singlet and keep the color of initial particles,
\textit{e.g.} Fig.~\ref{fig_scheme_a}. The calculation is well-known and under
theoretical
control~\cite{Bialas:1991wj,Schafer:1990fz,Khoze:2000cy,Bzdak:2004bx,Bzdak:2004ux}.
It can be performed using cutting rules or direct integration within the
kinematic regime where the momentum lost by the initial particles is small. 

However this simple model is not enough and to make a description more
realistic soft and higher order corrections need to be added, see
\cite{Kharzeev:2000jwa,Khoze:2000cy}. In the following we give a short
description of these corrections.

The impact factor~\cite{Martin:2001ms,Cudell:1995ki,Ivanov:2004ax}
regulates the infra-red divergence and embeds quarks inside the proton
as represented in Fig.~\ref{fig_scheme_b}. The impact factor is based
on a skewed unintegrated gluon density but its exact form depends on
the model considered.

The Sudakov form factor~\cite{Dokshitzer:1978hw,Khoze:2000mw,Coughlin:2009tr}
is one of the most important ingredients of the calculation. It corresponds to
virtual vertex correction (see Fig.  \ref{fig_scheme_c}) and depends on two
scales. The hard scale is linked to the hard subprocess ($gg \to X$). The soft
scale is related to the transverse momentum of the active gluons -- the scale
from which a virtual parton can be emitted. The Sudakov form factor suppresses
the cross section by a factor of the order of 100 to 1000.

Finally, additional pomeron exchanges between the initial and final state
protons can occur~\cite{Frankfurt:2006jp}, as schematically shown in
Fig.~\ref{fig_scheme_d}. This can lead to the production of additional
particles that might fill the gap created at the parton level. It is taken into
account by introducing the rapidity gap survival probability, which is a
probability of not having any additional soft interactions.

Each piece of the calculation can be investigated separately and its 
uncertainties can be estimated. The important point is that some of the
corrections are identical in all exclusive processes so that they can be
studied in one particular process and used to predict the cross section of any
process.

\subsection{The KMR Model}

The most quoted and first complete calculation is done in the Khoze, Martin and
Ryskin~(KMR) model from the Durham group. One can find here the main lines,
referring the reader to~\cite{Khoze:2000cy,Khoze:2000mw} for a review.

\begin{figure}[b] 
\centering
\subfigure[]{\includegraphics[width=4cm]{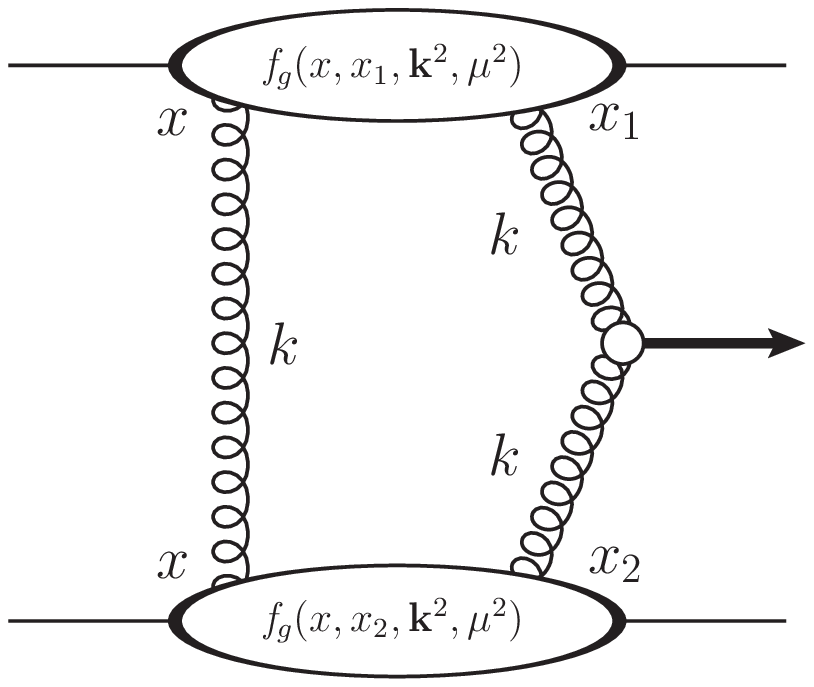}\label{fig_durhamCEP}} \quad
\subfigure[]{\includegraphics[width=4cm]{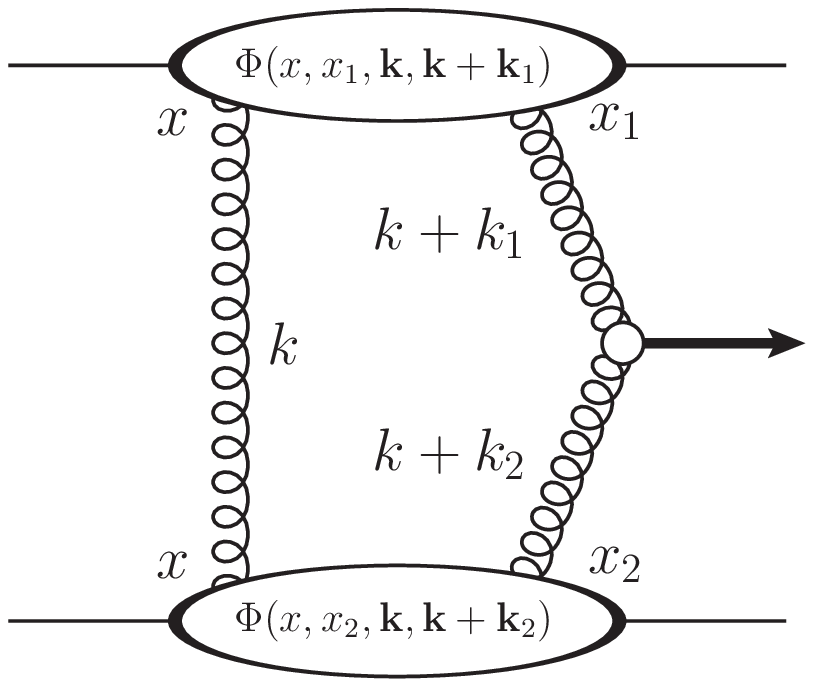}\label{fig_chideplot}} \quad
\caption{Schematic representation of the exclusive diffractive production
amplitude in the: (a) KMR model, (b) CHIDe model.}
\end{figure}

The cross section ($\sigma$) of the process represented schematically in
Fig.~\ref{fig_durhamCEP}, is assumed to factorize between the effective
luminosity~$\mathcal{L}$ and the hard subprocess~$\hat{\sigma}$:
\begin{linenomath} \begin{equation} \sigma=\mathcal{L}\times\hat{\sigma}(gg \to
X), \end{equation} \end{linenomath}
where $X$ is the centrally produced system. In particular
\begin{linenomath} \begin{equation} \frac{\partial\sigma}{\partial s\partial
y\partial{\bf P}^2\partial{\bf Q}^2}=S^2e^{-B\left({\bf P}^2+{\bf
Q^2}\right)}\frac{\partial\mathcal{L}}{\partial s\partial
y}\mathrm{d}\hat{\sigma}\left(gg\to H\right).  \end{equation} \end{linenomath}

The different variables are, the energy in the center-of-mass frame~$s$, the
rapidity of the centrally produced system~$y$ and the transverse momenta of the
final protons~${\bf P}^2$~and~${\bf Q}^2$. One can also recognize in turn, the
gap survival probability~$S^2$ and the~$t$-slope of the cross section
with~$B=4$~GeV$^2$ (taken from the fit to the soft hadronic
data~\cite{Khoze:2000cy}), introduced assuming that the dependence of the hard
cross section on the final proton transverse momentum is small. The subprocess
cross section for Higgs production, $\hat{\sigma}(gg \to H)$, includes an
additional~factor~$K$ fixed to~1.5, which takes into account
next-to-leading-order corrections. The effective luminosity is given by
\begin{linenomath}
\begin{multline} \label{eq_KMRlumi} \frac{\partial\mathcal{L}}{\partial
s\partial y}=\left(\frac{\pi}{\left(N^2_c-1\right)} \right. \\ \left.
\int\frac{\mathrm{d}\bk^2}{\bk^4}\mathit{f}_g\left(x,x_1,\bk^2,\mu^2\right)
\mathit{f}_g\left(x,x_2,\bk^2,\mu^2\right)\right)^2, \end{multline}
\end{linenomath}

$\mu$ is the hard scale and the variables are defined in
Fig.~\ref{fig_durhamCEP}.
The function~$\mathit{f}_g$ stands for the unintegrated skewed gluon density
related to the conventional integrated gluon distribution function and taken
here in their simplified form~\cite{Martin:2001ms}:
\begin{linenomath} \begin{equation} \mathit{f}_g\left(x,x_1\ll x,\bk^2,\mu^2\right)=
  R_g\,\frac{\partial}{\partial\log\bk^2}\left[\sqrt{
  T(\bk,\mu)}\,xg(x,\bk^2)\right].  \end{equation} \end{linenomath}

The factor~$R_g$ account for the skewness (the fact that~$x\neq x_1$;
$g(x,\bk^2)$ describes the forward gluon density only when~$x= x_1$) and is
found to be about~1.2 at the LHC energy of~14~TeV. One can note that the
Sudakov form factor~$T(\bk,\mu)$  
\begin{linenomath} \begin{multline} \label{eq_sudakov}
T(\bk,\mu)=\exp\left[-\int^{\mu^2}_{l^2}\frac{\mathrm{d}\bq^2}{\bq^2}
\frac{\alpha_s\left(\bq^2\right)}{2\pi} \right. \\ \left.
\int_0^{1-\Delta}\left(zP_{gg}(z)+\sum_qP_{qg}(z)\right)\mathrm{d}z\right],
\end{multline}\end{linenomath}
with~$\bq$ and~$z$ the transverse and longitudinal components of the additional
emission, is here included in the differentiation. $P_{gg}$ and $P_{qg}$ are
the quark and gluon splitting functions. In the KMR model, the presence of the
Sudakov form factor makes the integration infra-red stable and it is assumed to
provide applicability of perturbative~QCD.  According to a calculation at
single-log accuracy of the Durham group~\cite{Khoze:2000cy}:
\begin{linenomath} \begin{equation} \Delta=\frac{|\bq|}{|\bq|+\mu},\qquad
\mu=0.62\,m_X, \end{equation} \end{linenomath}
where $m_X$ is the mass of the centrally produced system. These results were
recently re-evaluated in~\cite{Coughlin:2009tr} giving new values for Higgs
production:
\begin{linenomath} \begin{equation} \label{eq_correction}
\Delta=\frac{|\bq|}{\mu},\qquad \mu=m_H.  \end{equation} \end{linenomath}
This correction leads to approximately a factor 2 suppression in the cross
section.

The KMR model has been developed for years and is one of the most complete
since it includes different types of exclusive diffractive production,~{\it
i.e.} from Higgs,~dijet,~$\gamma\gamma$, di-quark, $\chi_c$,~$\dots$~to
supersymmetric particles, and shows results in agreement with the available
data~\cite{HarlandLang:2010ep}.

\subsection{The CHIDe Model}

An other available model is the Cudell, Hern\'andez, Ivanov, Dechambre
exclusive (CHIDe) model~\cite{Cudell:2008gv} for jets and SM~Higgs boson
production. The structure of this model is similar to the one of the KMR model
but differs in the implementation and details of the different ingredients.  In
the CHIDe model the cross section for the exclusive process shown in
Fig.~\ref{fig_chideplot} is given by
\begin{linenomath}
\begin{multline} \sigma \simeq S^2 \Bigg[ \int
\frac{\text{d}^2\bk\,\text{d}^2\bk_1\,\text{d}^2\bk_2}{\bk^2(\bk+\bk_1)^2(\bk+\bk_2)^2}
\\[3mm] \Phi(x,x_1,\bk,\bk+\bk_1) \, \Phi(x,x_2,\bk,\bk+\bk_2)\\
\sqrt{T(\ell_1,\mu)}\,\mathcal{M}(gg\to X)\, \sqrt{T(\ell_2,\mu)}\Bigg]^2,
\end{multline} 
\end{linenomath}
where $\Phi$ is the impact factor, $T(\ell_i,\mu)$ is the Sudakov form factor,
$\mathcal{M}(gg\to X)$ is the hard subprocess amplitude, and the transverse
momenta $\bk$, $\bk_1$, $\bk_2$ are defined as in Fig.~\ref{fig_chideplot}. In
the whole calculation of the exclusive cross section, the exact transverse
kinematics is kept in all ingredients. Contrary to the KMR model the colour neutrality of the
proton is implemented independently of the Sudakov suppression in the impact
factor. It includes the skewed unintegrated gluon density and this
phenomenological model of the proton includes soft physics based on both the
data (elastic cross section, proton structure functions) and theory (dipole
picture, light-cone wavefunctions). It takes into account the proton
wavefunction as the impact factor goes to zero if one of the transverse momentum of
the~$t$-channel gluons goes to zero and the non-zero transverse momentum transfer is
introduced via a universal exponential factor.  The unintegrated gluon density
is built on the sum of two terms that take care respectively of the hard and
soft behavior of the proton structure function.  The hard component is based on
direct differentiation of the well-known gluon density
(GRV~\cite{Gluck:1998xa}, MRS~\cite{Martin:1998np} and CTEQ~\cite{Lai:1997vu}).
The soft component models soft colour-singlet exchanges in the non-perturbative
regime in the spirit of the dipole picture.  This gives
space for a contribution of the non-perturbative regime of~QCD. It was made in a
phenomenological way and therefore is not unique.  Actually, four different
fits are provided, all giving similar $\chi^2$ when adjusted to the~$F_2$ data~\cite{Ivanov:2000cm,Ivanov:2003iy}.
The main difference between the fits is the parametrisation of the soft
region~--~in particular, the transition scale from the soft to the hard regime.
They represent the present uncertainty on the unintegrated gluon distributions.

The Sudakov form factor is identical to~eq.~(\ref{eq_sudakov}). The upper limit
is taken at the Higgs mass according to the recent
result~\cite{Coughlin:2009tr} in the Higgs case but for dijet production it is
fixed to the hard-transverse momentum in the vertex.  Note that in this
model~$\Delta={|\bq|}/{\mu}$ and NLO corrections ($K$ factor) were also
introduced for the Higgs production. 

\subsection{Theoretical uncertainties}

The parton level computation is well understood and very precise.  However the
impact factor, Sudakov form factor and rapidity gap survival probability cannot
be calculated preturbatively and have to be modeled or parametrized. This
introduces non-negligible uncertainties that need to be discussed.

Three main sources of uncertainties can be identified concerning the
prediction of the exclusive jet or Higgs boson cross section. The
first one is the uncertainty on the gap survival probability. At $pp$
or $p \bar{p}$ colliders additional soft interactions can destroy the
gap in forward region or even the proton itself. While the Tevatron
measurement leads to a survival probability of 0.1, the value at the
LHC is still to be measured. We assume in the following a value of
0.03 at the LHC~\cite{Khoze:2000wk} and all mentioned cross sections
need to be corrected once the value of the survival probability has
been measured.

The two other sources of uncertainties and their effects on the
exclusive cross sections, namely the uncertainty on the unintegrated
gluon distribution in the proton and the constant terms in the Sudakov
form factor, will be discussed in the next paragraphs. In the CHIDe
model, the gluon density in the impact factor contains a soft and a
hard part. The hard part is known very well, mainly from the DIS
structure function~$F_2$ and vector meson data~\cite{Ivanov:2004ax},
but the soft one comes from a phenomenological parametrisation which
leads to uncertainties in both the dijet and the Higgs calculation.

In the dijet exclusive cross section, the main uncertainty comes
from the limits of the Sudakov integral, which have not yet been fixed by a
theoretical calculation. Therefore instead of eq.~(\ref{eq_sudakov}) in the
CHIDe model the following is used:
\begin{linenomath} \begin{multline} \label{eq_sudakov_CHIDe}
T(l_i,\mu)=\exp\Bigg[-\int\limits^{\mu^2/x}_{l_i^2/x'}\frac{\mathrm{d}\bq^2}{\bq^2}
\frac{\alpha_s\left(\bq^2\right)}{2\pi}\\ \int_0^{1-\Delta}\left(zP_{gg}+
\sum_qP_{qg}(z)\right)\mathrm{d}z\Bigg], \end{multline} \end{linenomath}
where two additional parameters, $x$ and $x'$, are included.  In the Higgs
exclusive case, the log structure of the Sudakov form factor has been
calculated to single-log accuracy and the complete one loop result can be taken
into account by adjusting the upper limit to~$\mu=m_H$, the lower limit is
$\bk+\bk_i$ with~$i$=1,2. However, this calculation does not take into account
the importance of the constant terms that cannot be exponentiated but
have important contributions when the coupling is running~\cite{Cudell:2008gv}.
To evaluate this theoretical uncertainty, we include the effect of changing the
constant terms by changing the lower scale. This gives a upper bound for the
uncertainty on the Higgs exclusive calculation because no uncertainty is
related to the lower scale itself~\cite{coughlin-forshaw}.

We discuss these uncertainties in  more detail
in Sec.~V, where we study the effect of varying these parameters and changing
the gluon densities. We also compare the results to the existing data.

\sidebyside{t}
{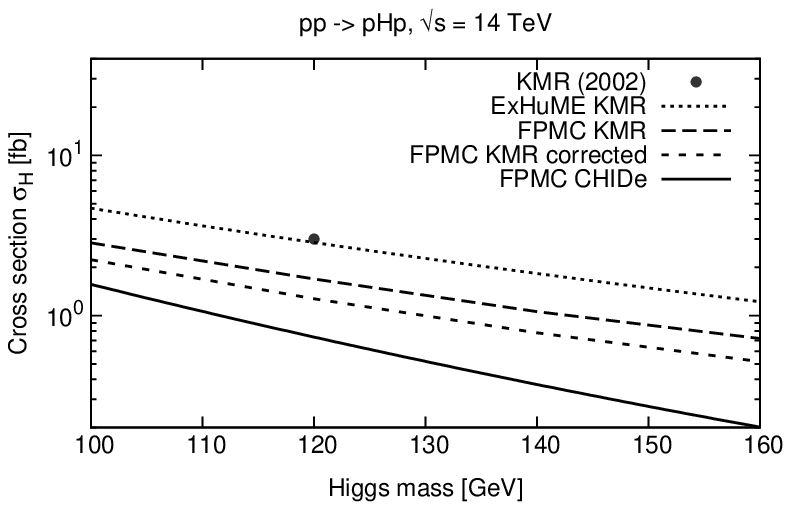}{models} {Cross section for exclusive Higgs boson production at the LHC as a
function of the Higgs boson mass. Predictions of CHIDe  and KMR implemented in
FPMC are presented. For comparison the implementations of the original KMR
model \cite{Khoze:2001xm} (black point) and ExHuME generator are given. In
addition the effect of changing the upper scale from $0.62m_H$ to $m_H$ in
eq.~(\ref{eq_correction}) on the KMR model is presented (FPMC KMR corrected). }
{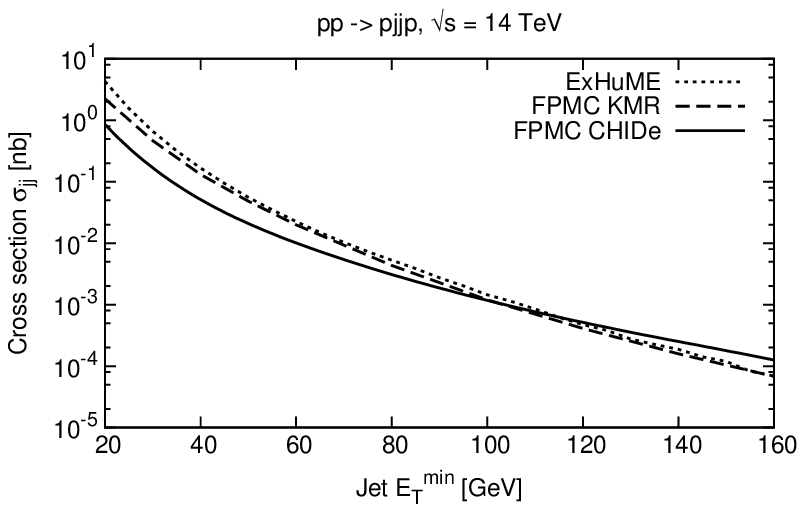}{models_jets} {Cross section for exclusive jet production at the LHC as a
function of of the minimum jet $E_T$. Predictions of CHIDe and KMR
are presented. For comparison the results of the ExHuME generator are given.}

\section{The Forward Physics Monte Carlo}

All models described above have been implemented in the  Forward Physics Monte
Carlo (FPMC) \cite{FPMC}, a generator that has been designed to study forward
physics, especially at the LHC. It aims to provide the user a variety of
diffractive processes in one common framework using HERWIG~\cite{HERWIG} for
hadronisation. In particular the following processes have been implemented in
FPMC: single diffraction, double pomeron exchange, central exclusive production
(including the direct implementation of KMR and CHIDe models) and two-photon
exchange (including anomalous couplings between gauge bosons).

The implementation of the KMR and CHIDe models in FPMC allows a direct
comparison of both models using the same framework. In Fig.~\ref{models}, we
display the cross section of exclusive Higgs boson production at the LHC for a
center-of-mass energy of 14 TeV as a function of the Higgs boson mass. In
addition, we display the predictions from the KMR original
calculation~\cite{Khoze:2001xm} and the results of the implementation of the
KMR model in the ExHuME generator~\cite{Monk:2005ji}. The difference in the
results between the FPMC and ExHuME implementations of the KMR model is the
effect of two factors. The first one is the different treatment of the gluon
distribution in eq.~(\ref{eq_KMRlumi}).  In ExHuME the value of the gluon
distribution is frozen for small $\bk^2$ (about \unit{1}{\GeV}), whereas in
FPMC we integrate from $\bk^2=\unit{2}{\GeV^2}$. In fact both solutions can
lead to uncertainties, therefore the better way is to introduce the modeling of
the soft region, which has been done in the CHIDe model.  The other reason of
the disagreement between FPMC and ExHuME is the different implementation of the
hard subprocess. In FPMC the Higgs is produced and then its decay is performed,
whereas the ExHuME implementation involves calculation of the Higgs propagator.
The difference on the Higgs production cross section between the KMR and CHIDe
models is clearly visible.  The CHIDe model leads to a smaller cross section
and shows a steeper dependence on the Higgs boson mass.  A similar difference
between models can be observed for the exclusive jet production at the LHC, see
Fig. \ref{models_jets}. The cross section obtained with the KMR model is higher
than the CHIDe prediction and a difference in slope is also visible. However,
as we will see in the following, these differences are within the uncertainties
of the models.

In order to compare the KMR and CHIDe models with the measurements performed by
the CDF Collaboration at the Tevatron the output of the FPMC generator was
interfaced with a cone jet algorithm of radius 0.7 as used by the CDF
Collaboration. 

\sidebyside{t}
{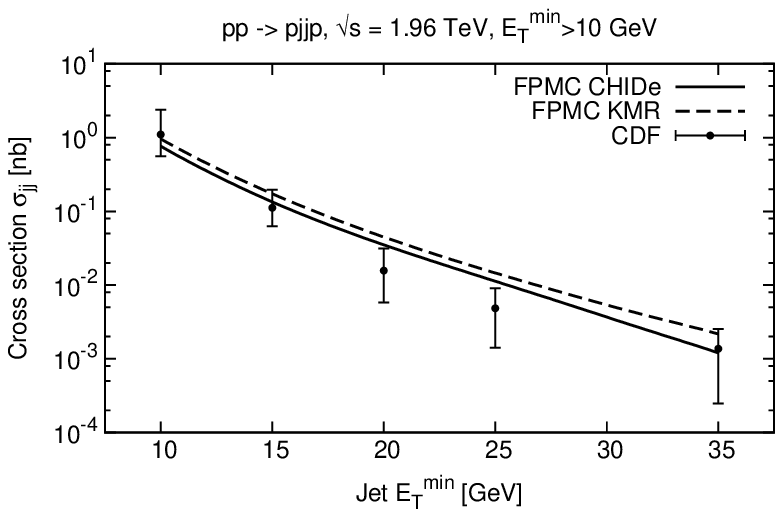}{CDF}
{Exclusive jet production cross section at the Tevatron as a function of the
minimum jet $E_T$. The CDF measurements are compared to the CHIDe and
KMR models displayed after applying the CDF jet algorithm.}
{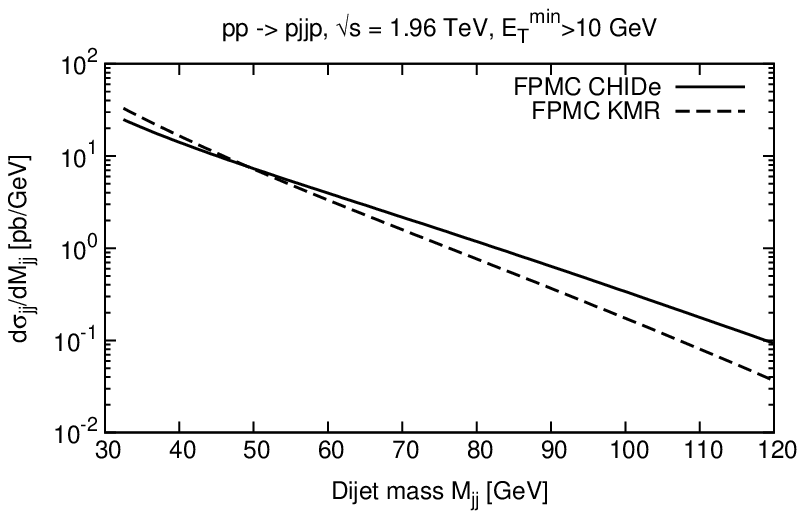}{Mjj_KMR_CHIDe}
{Dijet mass cross section for exclusive jet production at the Tevatron
for the CHIDe and KMR models.}

\sidebyside{t} 
{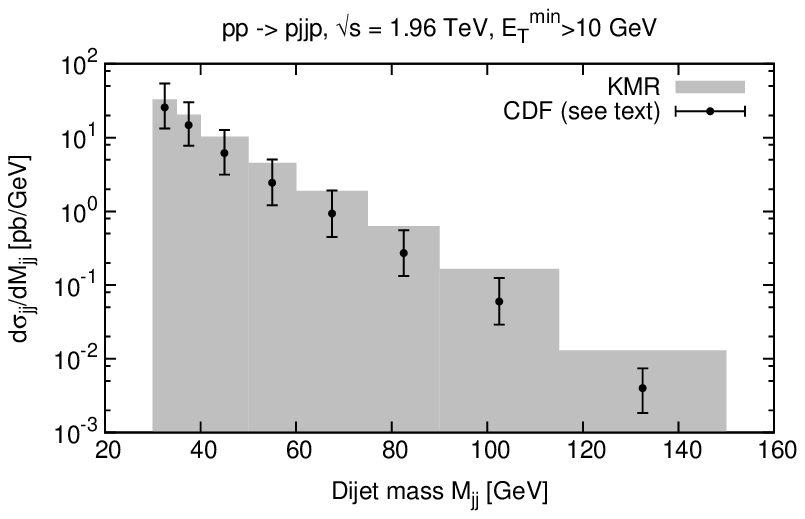}{Mjj_KMR_CDF}
{Dijet mass distribution extracted from the CDF measurement of exclusive jet
production
compared to the KMR model.}
{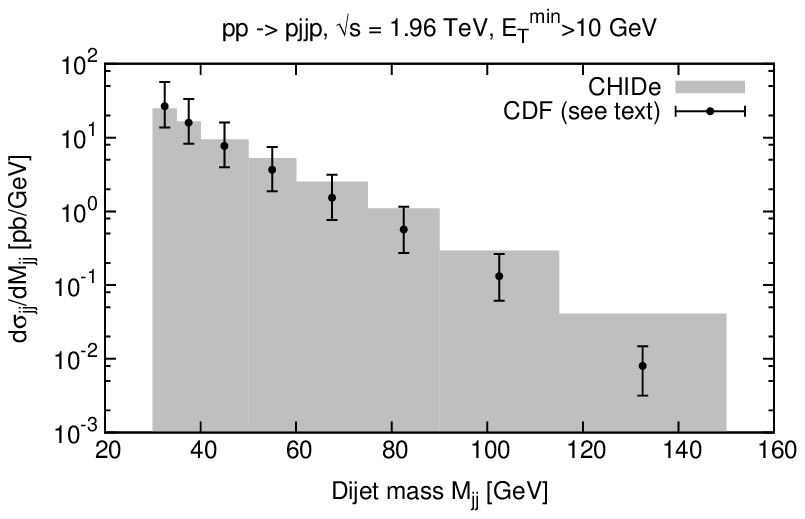}{Mjj_CHIDe_CDF}
{Dijet mass distribution extracted from the CDF measurement of exclusive jet
production compared to the CHIDe model.}

\section{Comparison to the CDF measurement}

To test the KMR and CHIDe models and their implementation in FPMC, the first
step is to compare their predictions with the measurements performed in the CDF
Collaboration at the Tevatron. The advantage of FPMC is that we can compare
directly the theoretical calculations with the CDF measurement since we use, at
the particle level, a 0.7 jet cone algorithm as used by the CDF Collaboration.
CDF measured the so called dijet mass fraction as a function of the jets
minimum transverse energy $E_T^{min}$ after tagging the antiproton in dedicated
roman pot detectors, and requesting a rapidity gap devoid of any activity in
the proton direction to ensure that only double pomeron exchange events are
selected. The dijet mass fraction is defined as the ratio of the dijet mass
divided by the total mass of the event computed using the calorimeter. If an
exclusive event is produced, it is expected that the dijet mass fraction will
be close to 1 since only two jets and nothing else are produced in the event.
On the contrary, inclusive diffractive events show some energy loss due to
pomeron remnants and the dijet mass fraction will be mainly distributed at
values lower than 1. The dijet mass fraction distribution allowed the CDF
Collaboration to separate the exclusive and inclusive diffractive contributions
and to measure the exclusive diffractive dijet cross section as a function of
the minimum jet $E_T$~\cite{Aaltonen:2007hs}.

The predictions of the KMR and CHIDe models are compared to the CDF measurement
in Fig.~\ref{CDF}. A good agreement is found between the CDF measurement and
the predictions of both CHIDe and KMR models and the difference between the
models is small compared to the data uncertainties. One should notice that the
data suggest slightly different dependence on $E_T^{min}$ that the models,
however it can just be a matter of statistical fluctuation.

Fig.~\ref{Mjj_KMR_CHIDe} displays the dijet mass ($M_{jj}$) distribution
predicted by the KMR and CHIDe models.  The difference in slope is very small, 
KMR leading to a slightly steeper dependence.

\sidebyside{t}
{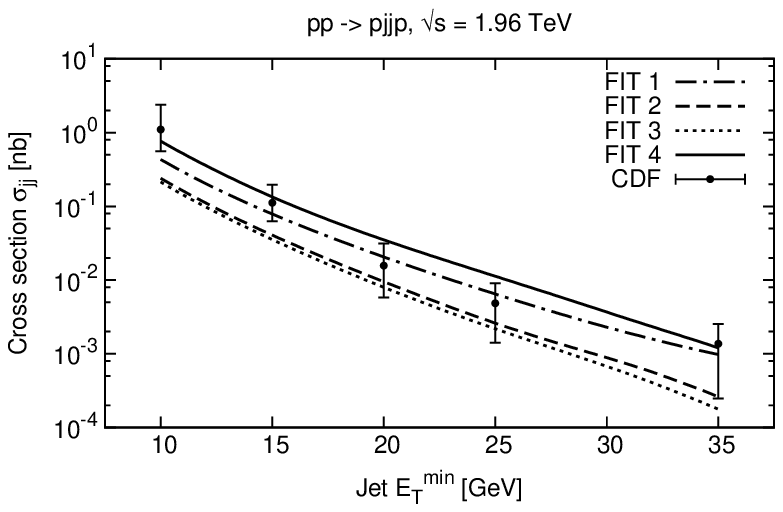}{GluonsTevatron}
{Effect of changing the gluon distribution on the exclusive jet production at
the Tevatron.}
{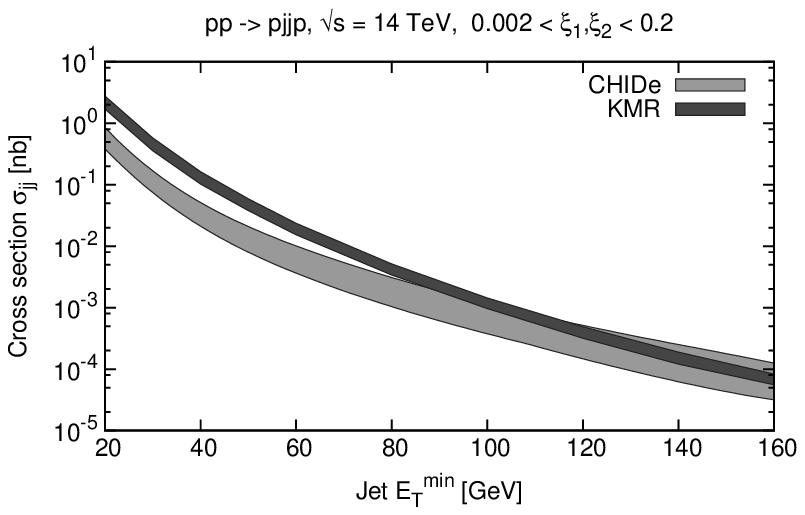}{GluonsLhcJets}
{Effect of changing the gluon distribution on the exclusive jet production at
the LHC.}

\sidebyside{t}
{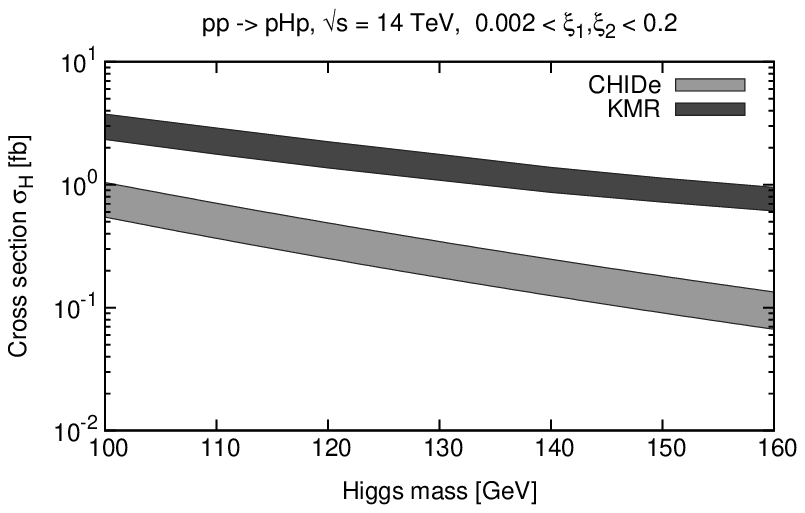}{GluonsLhcHiggs}
{Effect of changing the gluon distribution on the exclusive Higgs production at
the LHC.}
{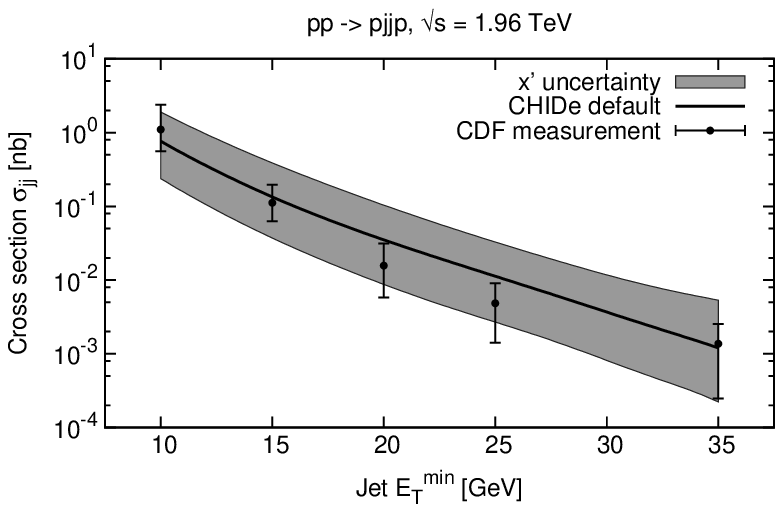}{LowerScaleTevatron}
{Effect of varying the lower limit of the Sudakov form factor on the exclusive
jets production at the Tevatron.}

In addition to the jet $E_T$ threshold dependence, the CDF Collaboration
published the exclusive jets cross section as a function of the dijet mass. The
dijet mass cross section is not a direct measurement but was extracted by the
CDF Collaboration from the jet $E_T$ threshold cross section data.  The method
is to compare the prediction of a given model (for instance KMR) with the
direct measurement of the jet $E_T$ threshold cross section. The MC predictions
are then reweighted to the CDF measurement in each  bin of $E_T^{min}$ (namely
10, 15, 20, 25, and 30 GeV) to obtain the CDF exclusive dijet mass cross
section ``measurement". The CDF ``measurement" can then be compared to the KMR
or CHIDe models. It is worth noticing that this method is clearly MC dependent
since there is not a direct correspondence between the jet $E_T$ and dijet mass
dependence. This is why we had to redo this study independently for each model,
namely KMR and CHIDe.

The comparisons between the CDF ``measurements" and the models predictions are
given in Figs.~\ref{Mjj_KMR_CDF} and \ref{Mjj_CHIDe_CDF} for the KMR and CHIDe
models, respectively. We stress once more that the CDF ``measurements",
displayed in both figures as black points, are model-dependent because of the
method used to extract them, and the ``data" points are different in both
figures. We note a good agreement between the CDF extracted measurements and
the KMR and CHIDe models displayed as gray histograms.

\section{Model uncertainties}

\sidebyside{t}
{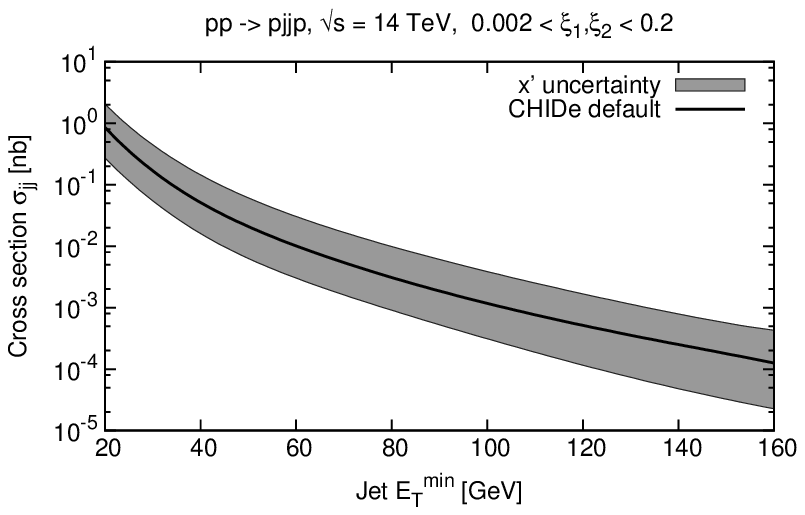}{LowerScaleLhcJets}
{Effect of varying the lower limit of the Sudakov form factor on the exclusive
jets production at the LHC.}
{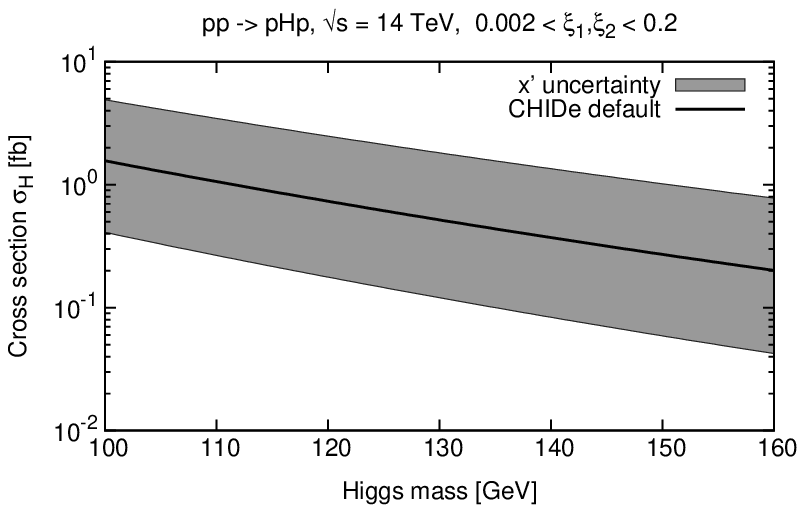}{LowerScaleLhcHiggs}
{Effect of varying the lower limit of the Sudakov form factor on the exclusive
Higgs production at the LHC.}

\sidebyside{t}
{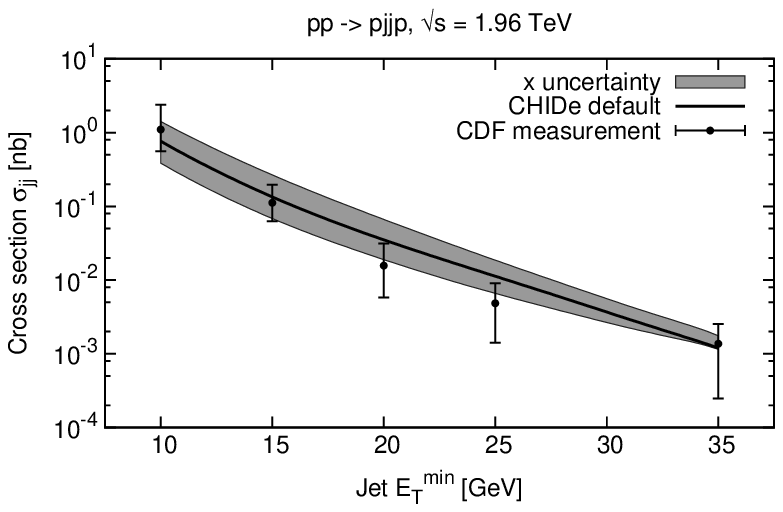}{UpperScaleTevatron}
{Effect of varying the upper limit of Sudakov form factor on the exclusive jets
production at the Tevatron.}
{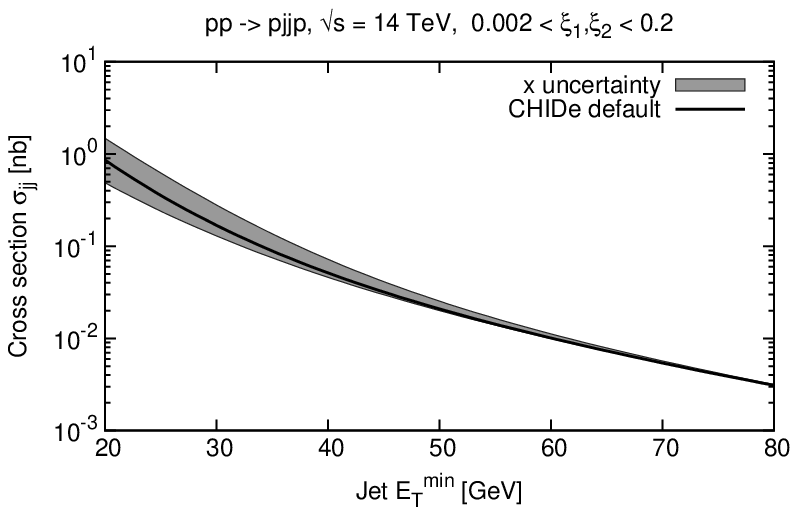}{UpperScaleLhcJets}
{Effect of varying the upper limit of the Sudakov form factor on the exclusive
jets production at the LHC.}

After having compared both KMR and CHIDe predictions to the present available
high-mass measurements from the Tevatron, we discuss in this section the
uncertainties of the model predictions especially for exclusive Higgs boson
production at the LHC.  In the following we discuss the uncertainties of the
CHIDe model.

To check the uncertainty due to the gluon distributions four different
parametrisations of unintegrated skewed gluon densities are used to compute the
exclusive jets and Higgs boson cross sections. As we mentioned in Sec. II,
these four gluon densities represent the uncertainty spread due to the present
knowledge on unintegrated parton distribution functions. The first step is to
check if these different unintegrated gluon distributions are compatible with
data.  Fig.~\ref{GluonsTevatron} shows the comparison between
the CDF measurement and the predictions of the CHIDe model using the four
different gluon distributions described above.  All gluon densities lead to a
fair agreement with the data. The measurement seems to favour FIT~4, but one
needs to remember that other parameters of the model, such as the cut off used
in the Sudakov form factor can modify the cross section as we will see in the
following.  There is an interplay between the different gluon distributions and
the scales used in the model. The default gluon density used in the CHIDe model
is FIT 4, which shows the highest soft contribution and predicts the highest
cross section.  Figs.~\ref{GluonsLhcJets} and \ref{GluonsLhcHiggs} show the
predictions of the CHIDe model with the same four gluon densities for the
exclusive dijet and exclusive Higgs at the LHC.  The uncertainty on the
exclusive cross sections due to the different gluon distributions is about a
factor of 3.5 for jets and 2 for Higgs boson, respectively. 

\sidebyside{t}
{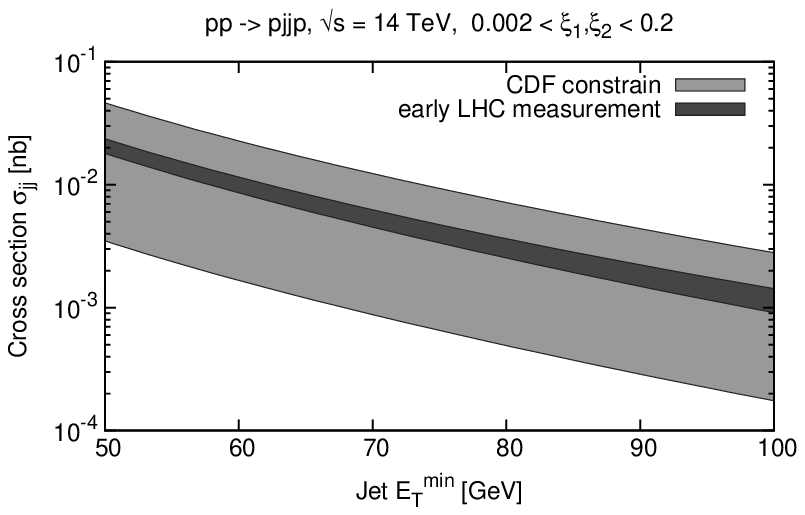}{LhcJetsGuess}
{Total uncertainty on the CHIDe model from the fit to the CDF measurement
(light gray) and possible exclusive jets  measurement with a low luminosity of
\unit{100}{\invpb} at the LHC (dark gray).}
{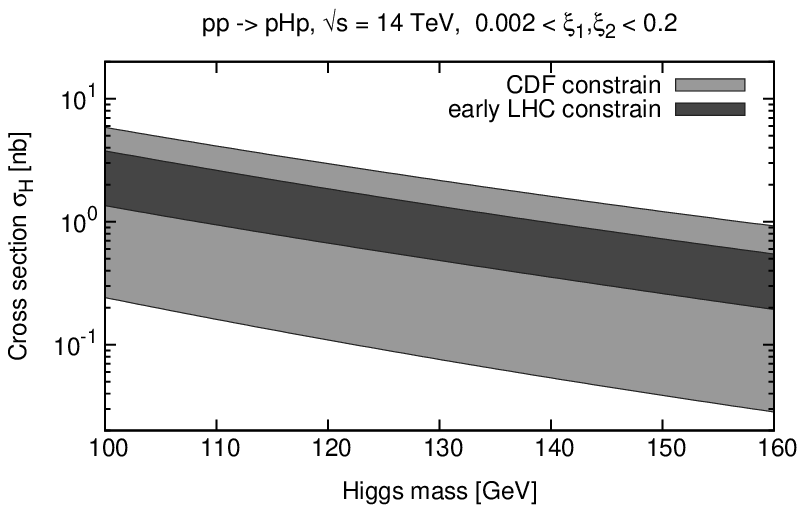}{LhcHiggsGuess}
{Total uncertainty on the CHIDe model for exclusive Higgs production at the
LHC: constraint from the fit to the CDF measurement (light gray), constraint
from possible early LHC jets measurements with \unit{100}{\invfb} (dark gray).}
                                                                               
For Higgs boson production at the LHC the uncertainty coming from the different
FITs is given by~Fig.~\ref{GluonsLhcHiggs}.  Using FPMC, one has the
possibility to study the uncertainty coming from the use of unintegrated gluon
density in models similar to the KMR model. This is done by changing the lower
cut-off in the unintegrated gluon. The bands showed in
Figs.~\ref{GluonsLhcJets} and \ref{GluonsLhcHiggs} correspond to a variation of
this cut-off on the gluon distribution from 1.26~GeV$^2$ (the minimal value at
which the gluon distribution MRST2002 is known) to 3~GeV$^2$. The difference is
small but not negligible.

In addition to the uncertainty due to the unintegrated gluon distribution, we
consider the additional uncertainties due to the values of integration limits
in the Sudakov form factor, see eq.~(\ref{eq_sudakov_CHIDe}).  Contrary to the
KMR model, the CHIDe model does not fix the limits of integration in the
Sudakov form factor in the dijet case. In the Higgs case, the upper scale is
fixed to $m_H$ and only the lower scale can be varied.
 
The lower integration limit is given by the $x'$ parameter and the default
value for FIT4 is 0.5, originally chosen \cite{DechambreThesis} to
describe the CDF data. Increasing the $x'$~value increases the values of the
integral and reduces the cross section. Decreasing $x'$ leads to the opposite.
Varying $x'$ by a reasonable factor of 2 up and down modifies the cross section
by a large factor up to 5 for all considered processes, namely jet production
at the Tevatron (see Fig.~\ref{LowerScaleTevatron}), jet production at the LHC
(see Fig.~\ref{LowerScaleLhcJets}) and Higgs boson production at the LHC (see
Fig.~\ref{LowerScaleLhcHiggs}).  

The upper limit of the integration is specified by the parameter $x$. As
already mentioned in section II, the value of the upper limit for the Higgs
boson production has been fixed by the calculation to 1.0 ($\mu = m_H$).
Although it still contributes to the total uncertainty for the jet production
cross section, its effect is much smaller for lower limit (see
Figs.~\ref{UpperScaleTevatron} and \ref{UpperScaleLhcJets} for the jet cross
section at the Tevatron and at the LHC respectively). The default value of the
$x$ parameter is 0.5, which we vary again by a factor 2. Decreasing its value
leads to an increase of the jet cross section. The effect is indeed visible at
Tevatron energies (Fig.~\ref{UpperScaleTevatron}) while it is negligible at the
LHC for $E_T^{min}$ above \unit{50}{\GeV} (Fig.~\ref{UpperScaleLhcJets}). It
should be noted that this is quite different from changing the $x'$ parameter.

From this analysis it follows that the uncertainty related to the exclusive
diffractive production is dominated by the uncertainty of the lower Sudakov
limit (that gives a rough estimation of the uncertainty coming from the
constant terms in the Sudakov form factor) for both jet and the Higgs boson
production. Also, at the LHC the uncertainty of the upper limit can be
neglected.  However this does not lead to a good estimation of the total
uncertainties on the Higgs cross section at the LHC. We need to check that the
variation of those parameters are compatible with the CDF measurement. In the
next section, we will study how to calculate the total uncertainty and how to
reduce it -- for the Higgs boson production -- using a possible early
measurement of exclusive jet at the LHC.

\section{Predictions for the LHC}

\begin{figure*}[t]
  \centering
  \includegraphics{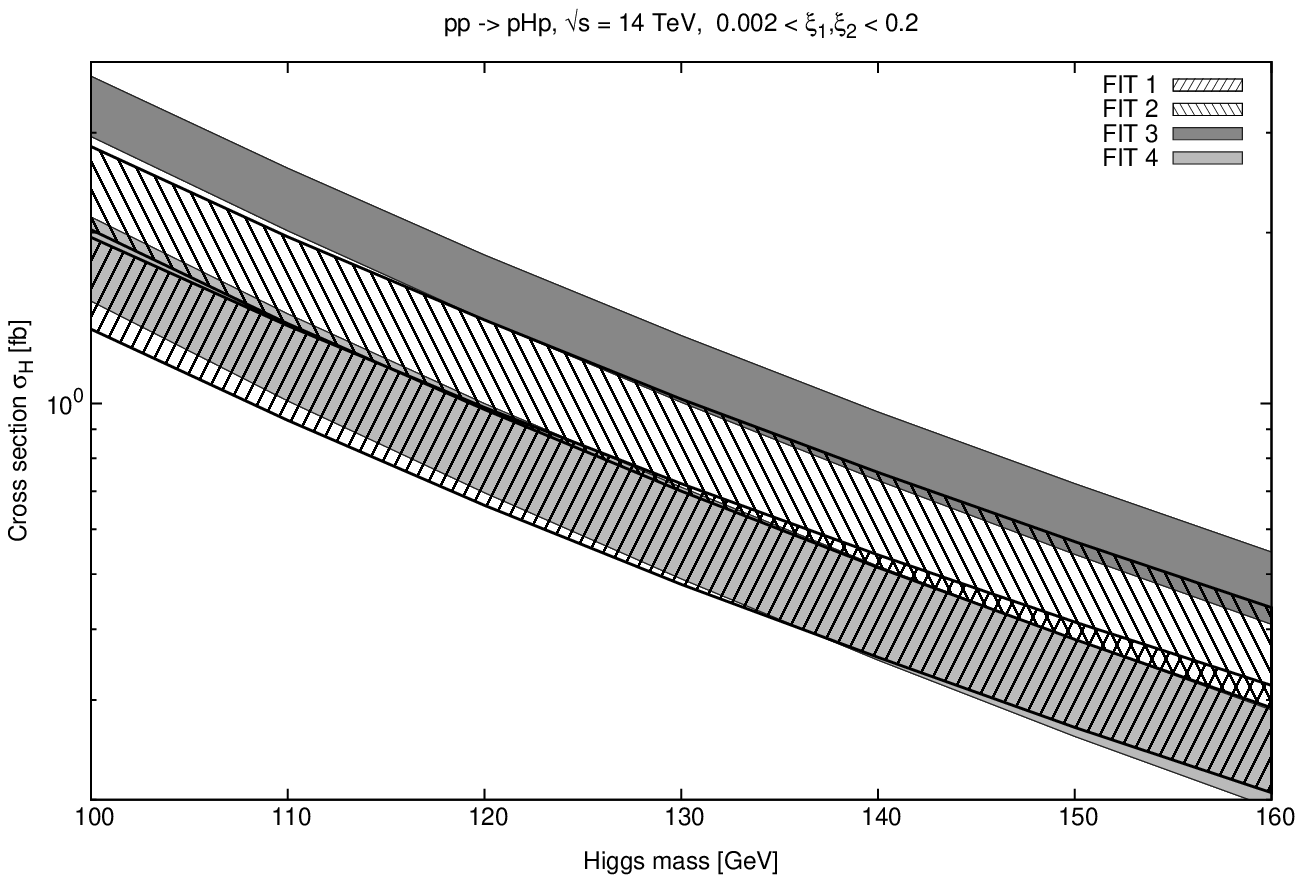}
  \caption{Contributions to the total uncertainty on the CHIDe model
    for exclusive Higgs production at the LHC. For each gluon density
    (FIT1 -- FIT4) the $x'$ uncertainty is shown for a luminosity of
    \unit{100}{\invfb}.}
  \label{LhcHiggsGuessGLU}
\end{figure*}

To make predictions for Higgs boson production at the LHC, we need to constrain
the model parameters using the Tevatron data.  The basic idea is to fit the
model parameters to the CDF measurement and extrapolate the model to the
center-of-mass energy of the LHC. We already know that the effect of the upper
limit of integration in the Sudakov form factor will be negligible for high
$E_T$ jets at the LHC compared to the effects from the lower limit and the
gluon density uncertainty. Varying the upper limit is not relevant for Higgs
production as we already mentioned in the previous section.

To study the impact of the uncertainties on the Higgs and jet cross sections
at the LHC, we need to take into consideration both the gluon uncertainty and
the lower limit of the Sudakov form factor calculation. The principle is
simple: for each gluon density (FIT1 to FIT4), we choose the $x'$ values which are
compatible with the CDF measurement to ensure that the model is indeed
compatible with Tevatron data for this given gluon density. Taking into account
the CDF data error, the procedure leads to two values of $x'$, namely
$x'_{min}$ and $x'_{max}$, for each gluon density. The same $x'$ values are
used at LHC energies to predict the jet and Higgs boson cross sections -- the
total uncertainty range is taken as the extreme values predicted by all gluon
densities, for appropriately $x'_{min}$ or $x'_{max}$.  The results are shown in
Figs.~\ref{LhcJetsGuess} and \ref{LhcHiggsGuess}.  The obtained uncertainty is
large, being greater than a factor of 10 for jets and about 25 for Higgs
production.

To study how the uncertainties on exclusive Higgs boson production can be
reduced, it is useful to check what the impact of the measurement of exclusive
jets at the LHC will be. This is quite relevant in order to reduce the present
uncertainty on the Higgs boson cross section.  We assume a possible early LHC
measurement of exclusive jets cross section for \unit{100}{\invpb}. In addition
to the statistical uncertainties, we consider a $3\%$ jet energy scale
uncertainty as the dominant contribution to the systematic uncertainty.  This
is quite conservative but takes into account other sources of uncertainties such
as jet energy resolution and we assume this measurement to be performed at the
beginning of the data taking of the LHC when all detectors are not yet fully
understood.  A possible result of such measurement is presented in
Fig.~\ref{LhcJetsGuess}.  It is clearly visible that even very early LHC data
can constrain the models much more than the Tevatron.

To check how this new measurement can constrain further the model
uncertainties, we follow the same procedure as in the beginning of this section
when we used the CDF data.  For each gluon density, a range in $x'$ describing
the exclusive jets measurement at the LHC is chosen. The LHC early measurement
can constrain the uncertainty on the Higgs boson production cross section to
about a factor 5, as shown in Fig.~\ref{LhcHiggsGuess}.  The exclusive jet
cross section measurement at the LHC allows to constrain the $x'$ parameter for
each gluon distribution. Fig.~\ref{LhcHiggsGuessGLU} shows the uncertainty of
each gluon density separately. Although the uncertainty caused by the $x'$
values is small, the remaining uncertainty due to the different gluon densities
is large. Therefore some other measurements such as the exclusive photon
production are needed to constrain further the Higgs cross section at the
LHC.

\section{Conclusions} 

The KMR and CHIDe models of the exclusive jets and Higgs production have been
implemented in the FPMC generator. They both show very similar, good
description of exclusive jets measurement at the Tevatron energy. Although the
predictions for the LHC energy show large differences, they are within the
uncertainties of the models.

The main sources of uncertainties at LHC energies are the uncertainties on the
gluon density in the soft region and the Sudakov form factor.  Taking them into
account, the results of the KMR and CHIDe models are compatible. The total
uncertainty estimated from the CHIDe model predictions is quite large -- a
factor 10 for jets and 25 for Higgs, after taking into account the constraint
coming from the CDF exclusive jets measurement. Further measurements at the
Tevatron ($\chi_c$ or exclusive photons) will constrain the model further. We
are getting the upper bound of the uncertainties that can be greatly reduced
when measurements of the exclusive jets at the LHC are available. An early
measurement using \unit{100}{\invpb} can constrain the Higgs production cross
section by a factor of 5.

\section{Acknowledgments}
A.D. would like to thank J.-R. Cudell and I. P. Ivanov for discussions and
corrections as well as the CEA-Saclay for its support.

\end{document}